\documentclass{icrc29}
\usepackage{graphicx,amssymb,amsmath,times}
\begin{document}
%Title of paper
\title[High-Energy Neutrino Astronomy with the Super-Kamiokande Detector]{High-Energy Neutrino Astronomy with the Super-Kamiokande Detector}
\author[A. Habig et al.] {A. Habig$^a$,
  for the Super-Kamiokande Collaboration\\
        (a) Univ. of Minnesota Duluth Physics Dept., 10 University Dr.,
        Duluth, MN 55812, USA\\
        }
\presenter{Presenter: A. Habig (ahabig@umn.edu), \  
usa-habig-A-abs3-he23-poster}

\maketitle

\begin{abstract}

The Super-Kamiokande experiment has collected a large sample of
high-energy neutrino events.  These are primarily atmospheric neutrinos,
but a bright enough astrophysical source could also be visible.  The
data have been examined for possible point and bursting neutrino
sources, as well as possible WIMP annihilation signatures.  No
significant evidence for such sources have been found, and the resulting
flux limits have been calculated.

\end{abstract}

\section{Introduction} 

The highest energy neutrinos observed in the Super-Kamiokande experiment
are seen via the upward-going muons which enter the detector when the
neutrino interacts in the rock surrounding the experiment.  Those muons
which have enough energy to pass through the whole detector are called
``through-going'' and come from parent neutrinos with a typical energy
of 100~GeV.  Those which stop in the detector (``stopping'') are made by
neutrinos with typical energies of 10~GeV.  More details of the data and
their use in the analysis of neutrino oscillations can be found
in~\cite{SK-full}.  An additional ``showering'' subset of the data with
typical neutrino energy of 1~TeV has recently been identified by
selecting upward-going muon events that experience radiative energy
losses~\cite{icrc-shantanu}.  Also, an even higher energy sample has
been recovered from extremely bright muons which saturate the Inner
Detector and are bypassed entirely by the standard data reduction
process.

The high energy end of the observed Super-K neutrino spectrum is of
astronomical interest due to the steeply falling atmospheric neutrino
spectrum.  While neutrinos of energy greater than a GeV will produce
leptons which follow the parent neutrino direction reasonably well, thus
allowing one to identify where that neutrino came from on the sky, below
a TeV the known atmospheric neutrino flux is much greater than predicted
astrophysical neutrino fluxes~\cite{learned-mannheim}.  The higher
energy sample one can study, the better chance one has of picking out an
astrophysical neutrino signal above the atmospheric neutrino background,
and as a bonus the higher the parent neutrino energy the more closely
the resulting muon follows the initial neutrino direction.  Similar
searches have been performed in the past.  This paper presents new
results from the same dataset as Super-K last
presented~\cite{icrc-kristine}.  A thorough explanation of some the
general methods used was written by MACRO~\cite{macro-nuastro}, and the
AMANDA experiment is rapidly collecting a large dataset covering the
northern sky~\cite{amanda-nuastro}.

\section{Astrophysical Neutrino Searches with Super-K}

The neutrino data come from the first (pre-implosion) phase of the
Super-Kamiokande experiment, ``SK-I'', April 1996 through July 2001, a
live-time of 1680 days.  1892 through-going (including 309 high-energy
``showering'' events) and 467 stopping muons were observed.  Previous
work~\cite{icrc-kristine} looked for any DC excesses indicating point
sources, a correlation with the 1997 Mrk~501 outburst, and set DC flux
limits around known high-energy sources.  Super-K has also searched for
neutrinos in coincidence with GRB's~\cite{sk-grb} and from WIMP
annihilation's in astrophysical gravitational potential
wells~\cite{sk-wimp}.  No sources have yet been found.

\subsection{Soft Gamma Ray Repeaters}

Soft Gamma Ray Repeaters (``SGRs'') are a small class of bursting high
energy sources.  Unlike classic gamma ray bursts, they repeat, have a
softer gamma spectrum, and are located in the galactic
plane~\cite{sgr-obs}.  ``Magnetar'' models predict they might produce a
neutrino flux, albeit a smaller one than to which Super-K is
sensitive~\cite{sgr-nus}.  IPN data~\cite{ipn} was used to identify
bursts from four known SGRs.  Upward-going muons within 74 different
windows of $\Delta T = \pm 1$~day and $\Delta \theta = 5^{\circ}$ around
these bursts were selected, and one event was found.  The background of
random coincidences was found to be 0.013 events per window.  With a
trials factor of 74, the total expected background is 0.96 events,
consistent with the one observed, so no evidence for neutrinos from SGRs
was found~\cite{shantanu-thesis}.

\subsection{Untriggered Burst Search}

Extending the searches for GRBs, SGRs, and Mrk~501 outbursts to a more
general case, an untriggered all-sky burst search was done.  The
aforementioned searches are examples of triggered burst searches -- that
is, the time and place of the astrophysical burst are used to look for
correlations in the SK neutrino signal.  To be free of trigger bias,
most notably to have a chance of picking up neutrino bursts that for
whatever reason were not seen by high energy electromagnetic telescopes,
the data were examined for any self-correlation between upward-going
neutrino-induced muons.

The method used was to regard each upward-going muon as a ``trigger''
itself, and to check for other such events arriving within an hour and
$5^o$ on the sky.  This is similar to the neutrino multiplicity analysis
presented in~\cite{macro-nuastro,icrc-kristine} with an additional time
cut.  One such doublet was found in the SK-I data.  The expected
background is $2\times10^{-5}$ such coincidences, but when multiplied by
the trials factors (number of upward-going muons minus one), the total
chance of seeing such a chance coincidence is 5\%, not statistically
significant albeit tempting.

\subsection{Searches with the Showering muon data subset}

\begin{figure}[h]
  \begin{center}
    \begin{minipage}[c]{0.4\textwidth}
      \includegraphics[width=\textwidth]{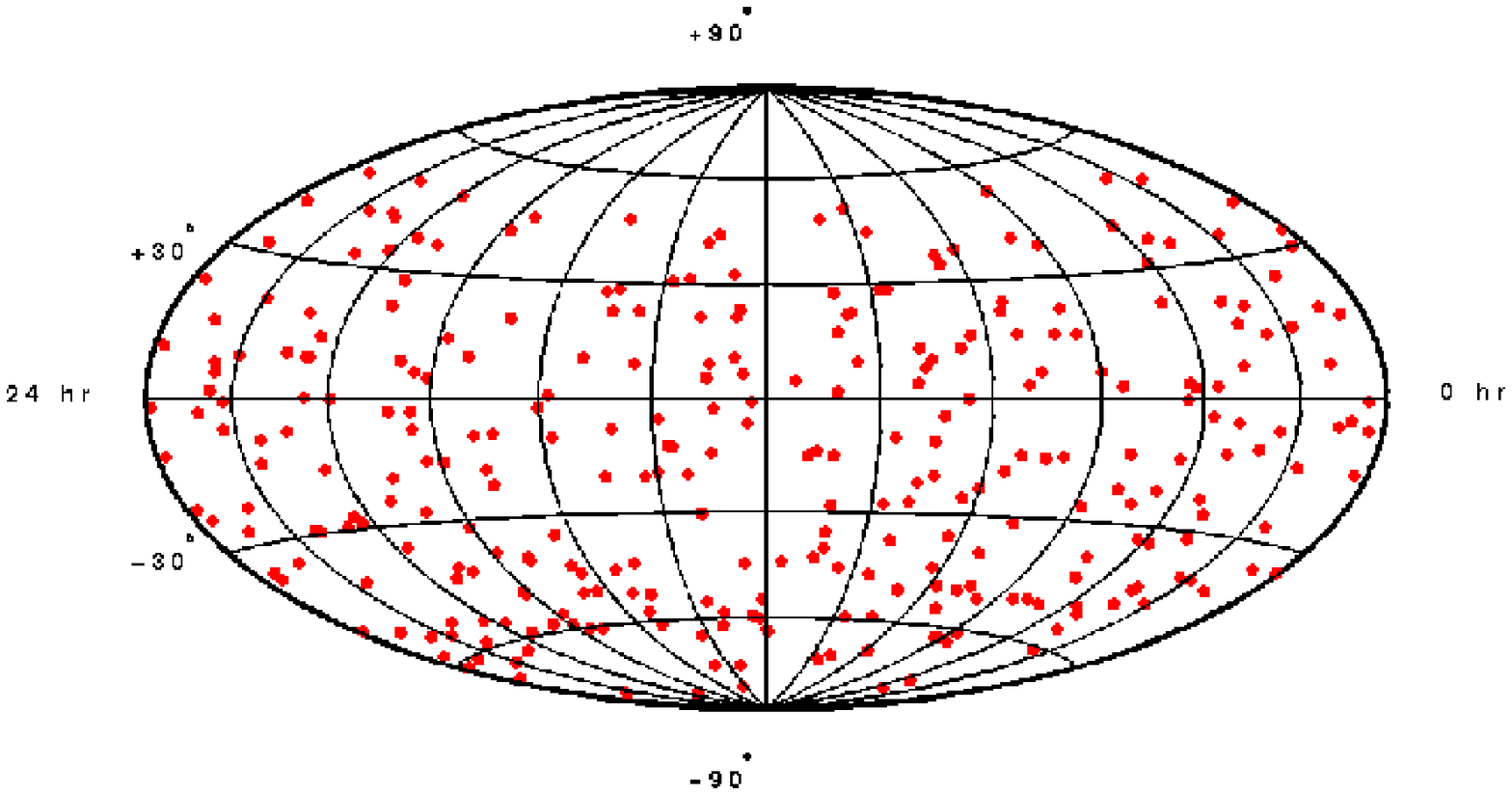}
    \end{minipage}
    \begin{minipage}[c]{0.4\textwidth}
      \includegraphics[width=\textwidth]{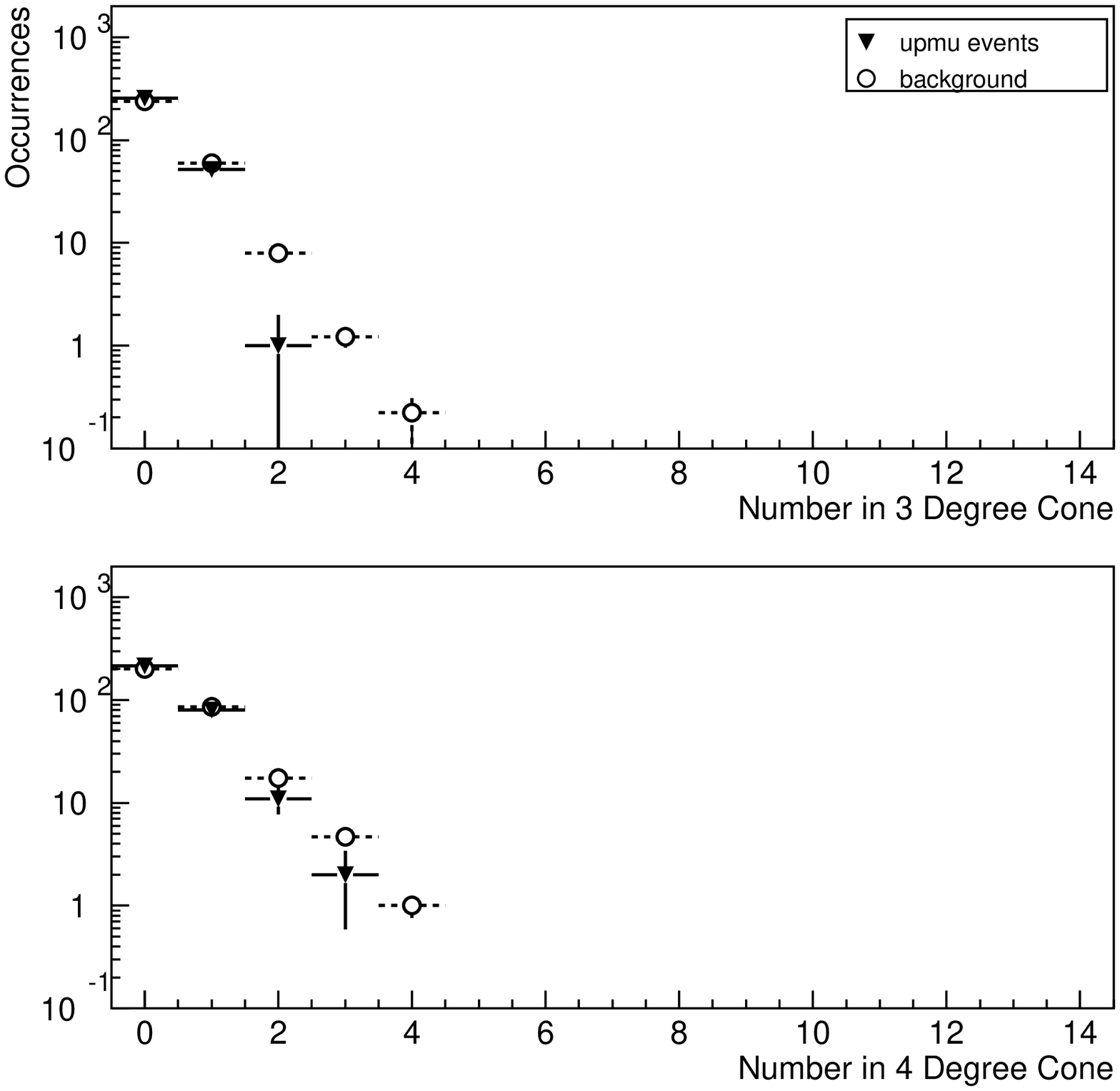}
    \end{minipage}

    \caption{\label{fig:showeringfigs} 
      On the left is a map in Equatorial Coordinates of the high-energy
      ``showering'' subset of Super-K's upward-going muon data.  These
      events show no evidence of unusual clustering which would
      indicate a possible point-source, as seen in the plots on the
      right.  These graphs show the frequency of muon coincidences as a
      function of multiplicity, plotting the number of other events
      within $3^\circ$ and $4^\circ$ about each event.  The triangles
      are the data, the circles from the atmospheric neutrino Monte
      Carlo and show the expected degree of coincidental clustering.
    }

  \end{center}
\end{figure}

The higher neutrinos in the showering data sample offer a better chance
to beat the soft background of atmospheric neutrinos.  Thus, many of the
same searches presented previously which used all upward-going muons
have been re-done with just this high energy sample.  These searches
are: a search for DC excesses around known high-energy astrophysical
objects; an all-sky search by checking for clustering of events in
space; a search for an excess of events coming from the center of the
Earth, Sun, and Galaxy to probe for WIMP annihilation; and a search for
an excess of such events coming from the galactic plane, which probes
for cosmic ray interactions in the interstellar medium.  No
statistically significant excesses were found.
Fig.~\ref{fig:showeringfigs} shows the all-sky map of such events and
the distribution of cluster multiplicities compared to the expected.

\subsection{Diffuse Flux Limits with the Highest Energy Upward-going Muons [preliminary]}

The very highest energy muons observed in Super-K are directed to a
separate analysis chain, as nearly all the photomultiplier tubes
(``PMTs'') have been saturated, causing problems for the standard
analysis tools.  52,214 events of more than $1.75\times10^6$
photoelectrons were observed in SK-I.  As with more sedate muons, nearly
all of these events are downward-going cosmic ray muons.  Using the
timing and charge of the Outer Detector veto shield PMTs, which collect
less light and thus remain unsaturated, the directionality of these
high-energy muons was checked.  A result of one upward-going
muon was found.

\begin{figure}[h]
  \begin{center}
    \includegraphics[width=0.8\textwidth]{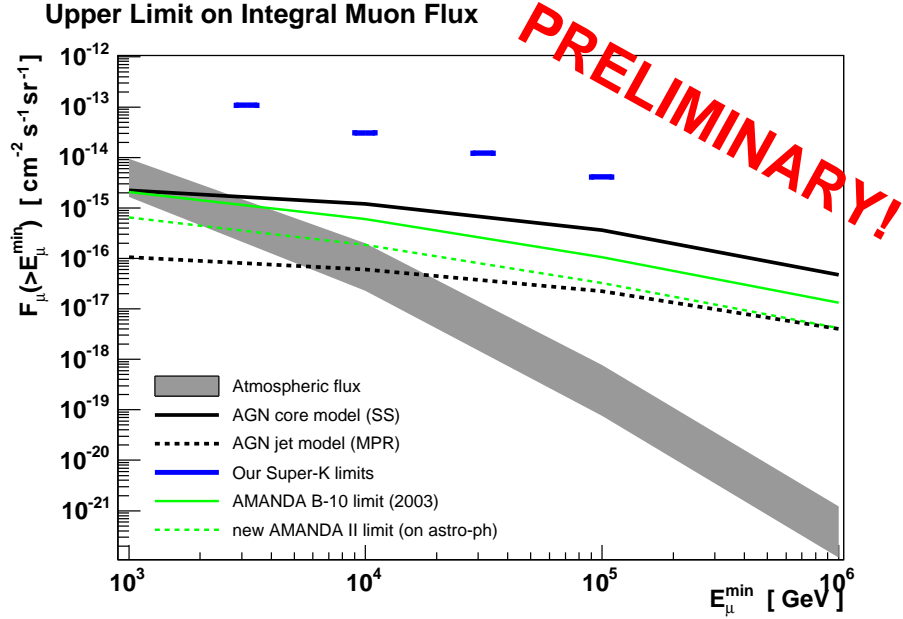}

    \caption{\label{fig:diffuselimit} 
      Given the observation of one upward-going muon which deposited
      more than $1.75\times 10^6$ photoelectrons in Super-K compared to
      the expectation of $0.47\pm0.25$ events at these energies from
      atmospheric neutrinos, limits can be set on the muon flux induced
      by very high energy neutrinos coming from astrophysical soruces.
      The upper limits from Super-K (blue dashes) are shown as muon flux
      above a threshold muon energy $E_\mu^{min}$. These limits are
      compared to muon flux inferred from the AMANDA experiment's
      neutrino flux limits (green lines)~\cite{amanda-diffuse}, the
      expected muon flux due to atmospheric neutrinos (shaded
      region)%~\cite{candiaflux} 
      and representative models of muon flux
      due to neutrinos from AGNs (black lines)~\cite{szabo,stecker}.
%      Muon fluxes were determined using neutrino cross sections from
%      CTEQ3-DIS~\cite{cteq} and the Lipari-Stanev effective muon
%      range~\cite{lipari}.
      }

  \end{center}
\end{figure}

Detection efficiencies of muons with energies greater than 3 TeV were
determined using Monte Carlo simulations.  Given the small number
statistics involved, binning the data on the sky is not feasible, so the
data was compared against a prediction for the whole sky given the known
atmospheric neutrino spectrum.  An expected background of $0.47\pm0.25$
atmospheric neutrino-induced extremely bright muons was found.  Given
the one observed event, an upper limit as a function of energy has been
calculated for neutrino-induced muons of energies 3--100~TeV from
possible astrophysical sources, to compare to similar searches from
%MACRO~\cite{macro-diffuse} and 
AMANDA~\cite{amanda-diffuse}.  These
limits are shown in Fig.~\ref{fig:diffuselimit}.

\section{Conclusions}

In an effort to identify possible astrophysical neutrinos, the SK-I
dataset was examined in new ways to extract the highest energy neutrinos
available to this detector.  Showering muons exhibit radiative energy
loss and come from a typical parent neutrino energy of 1~TeV.  No
statistically significant excess of these neutrinos was observed.  The
very highest energy muons in Super-K were recovered from the saturated
PMT data.  One was found to be upward-going, consistent with the
expectations from the atmospheric neutrino spectrum.  A search of all
upward-going neutrinos compared to SGR outbursts also yielded no
significant sign of an astrophysical source, as did an untriggered burst
search checking for event clusters in time and space.

\section{Acknowledgments}

We gratefully acknowledge the cooperation of the Kamioka Mining and
Smelting Company.  The Super-Kamiokande experiment has been built and
operated from funding by the Japanese Ministry of Education, Culture,
Sports, Science and Technology, the United States Department of Energy,
and the U.S. National Science Foundation.  The bulk of the analysis
presented in this paper was done by Shantanu Desai (now at Penn State)
and Molly Swanson (MIT).  This presentation was directly supported by
NSF RUI grant \#0354848.

\end{document}